# 3rd grade English language learners making sense of sound

Enrique Suarez and Valerie Otero

*School of Education, University of Colorado, Boulder, 80309, USA*

**Abstract.** Despite the extensive body of research that supports scientific inquiry and argumentation as cornerstones of physics learning, these strategies continue to be virtually absent in most classrooms, especially those that involve students who are learning English as a second language. This study presents results from an investigation of 3rd grade students' discourse about how length and tension affect the sound produced by a string. These students came from a variety of language backgrounds, and all were learning English as a second language. Our results demonstrate varying levels, and uses, of experiential, imaginative, and mechanistic reasoning strategies. Using specific examples from students' discourse, we will demonstrate some of the productive aspects of working within multiple language frameworks for making sense of physics. Conjectures will be made about how to utilize physics as a context for English Language Learners to further conceptual understanding, while developing their competence in the English language.

## INTRODUCTION

With the expected shifts in our country's population, schools will see an increase in the number of students who come from non-English speaking backgrounds; a large number of these students are themselves learning English as a second language [1]. These demographic changes have understandably prompted research on second-language acquisition, as well as the design of Language Arts curricula that help these English Language Learners (ELLs) develop language skills. Nevertheless, a search in ERIC (www.eric.ed.gov) shows that there have been fewer efforts that attempt to understand how ELLs recruit and expand everyday language in service of making sense of science as it is presented in school. Understanding the role that language plays in how ELLs learn science is particularly important, especially when it is assumed that science learning is mediated by social interactions. Furthermore, evidence-based classroom science is particularly well suited for helping students communicate in English around shared classroom experiences. This study explores how emergent learning environments, where students can use familiar language to think about and explain observations, promote the development of language skills and conceptual understanding. More specifically, we address the following questions: (i) how do ELLs use everyday language in the service of understanding physics and (ii) how do everyday and academic language interact with each other during the students' meaning-making process and concept formation?

In order to address these questions, we analyze a discussion between third grade ELLs, in which they tried to establish a connection between physical properties of strings (length, tension, frequency of vibration) and the characteristics of the sounds produced by them (pitch and volume).

## RESEARCH CONTEXT

Thirteen third grade students participated in this study. These students were enrolled in large K-8 urban public school that ran two separate academic programs: a "Mainstream" program for monolingual English speakers and students considered proficient in English; and a Sheltered English Immersion Program (SEIP) for students who were learning English as a second language. See Table 1 for demographic breakdown.

**Table 1.** Demographic composition of the School

| ESL | Free & Reduced lunch | Hispanic | White | Asian | African American |
|---|---|---|---|---|---|
| 66% | 76% | 45% | 31% | 13% | 9% |

Data were collected from a beginner/intermediate SEIP classroom, where nine different first languages were spoken, and students and their families came from nine different countries. Students' length of residence in the country ranged from US-born, to arriving up to three months before recording the session we present below.

The episode we analyze in this study occurred during the science unit of Sound. In the previous sessions, the teacher and the students had reviewed the academic concept of vibrations; made connections to students' experiences with vibrations; defined four characteristics of vibrations ("volume, pitch, speed, and size") and introduced academic terminology for referring to them; performed an experiment in which students flicked a ruler and recorded their observations of the four characteristics mentioned above; and had started hypothesizing about the link between the length of the part of the ruler extending from the edge of the desk and the frequency and size of the vibrations.

## THEORETICAL FRAMEWORK

We assume that students' development of scientific knowledge is co-constructed through social interactions and cultural practices; therefore, relying heavily on language and literacy [2]. For ELLs, these communication expectations involve a different set of challenges than for native English speakers, mainly due to incompatibilities with students' familiar registers, English language discourse patterns, and the academic discourse practices of a discipline. This incongruence in familiar communicative practices and those presented through schooling can lead to the disfranchisement of students from any group and could greatly hinder their learning process [3].

Warren *et al*. [3] suggest that, when the learning environment allows for it, ELLs, like most students, resort to everyday language when communicating their ideas about physical phenomena. Classrooms that invite familiar registers and practices into science activities, encourage students to engage in productive sense-making, reasoning mechanistically [4], and argumentation; all crucial for developing conceptual understanding in science. Moreover, by presenting ideas and actively participating in discussions, students continue to develop and improve English language skills.

In order to better understand the affordances of classrooms that recognize and value students' everyday language, we study these environments through the framework of *Third Space* [5].

### Third Space

The current educational system tends to privilege academic language introduced and used through schooling, and often discourages the use of everyday language associated with out-of-school spaces. The perspective of "Third Space" is drawn from the work of Vygotsky [6], which contrasts the mediating role of everyday and academic language, and practices, through schooling. Third space considers students' community as Space 1, where common and familiar registers are used; and the school classroom as Space 2, where technical/academic vocabulary is favored during teaching and learning. In the particular case of literacy, Gutierrez *et al*. [5] suggest that literacy *development* is related to students' deployment of everyday language in the service of their trying and testing of formal literacy practices. At the same time, students' everyday language are generalized as students' apply them to the formal uses of language that are presented through schooling. Through this bi-directional process, students construct meaning in the context of formal schooling [7]. The "space" in the classroom in which formal and everyday language and ways of knowing interact defines an alternative, emergent Third Space that bridges everyday and academic spaces, "creating the potential for authentic interaction and learning to occur" [3, p. 372] These are pedagogical arenas where students can use and rely on their culturally mediated registers when testing out, and communicating, ideas and knowledge associated with formal terminology and classroom practices. Physics is particularly suited for instruction that capitalizes on this model of learning because of its appeal to shared experiences with everyday, observable phenomena.

Physics activities that invite students to use everyday language in conjunction with academic language help students in the processes of English language development, as well as in the construction of scientific language and meaning.

## METHODOLOGY

To investigate our research questions, we videotaped activities in which students explored and discussed their ideas about sound, with video footage from four Sound unit lessons. The episode analyzed below occurred during the third session; it shows a discussion between students, trying to understand how the physical properties of strings on a guitar-like instrument were connected to the characteristics of the sound they heard. The instrument built by the teacher consisted of a pegboard with hooked screws on one end, and fishing line tied on one end to a screw and the other end to one of the holes on the board.

Guided by literature on children's conceptions of sound [8], we used a generative coding methodology to determine the everyday and academic language used by students, and mechanisms they identified when talking about the sounds they heard and the connection to the instrument. We then established three coding categories based on students references to physical properties of strings: length, tension*,* and vibration frequency. These codes were then further investigated in terms of how, and for what purposes, students used specific everyday and academic terminology.

# FINDINGS

We found that students recruited familiar language for talking about the different sounds they heard. Some of this everyday language was taken up by the class and played an important role in students' negotiations about the mechanisms that drive differences in pitch. There is also evidence suggesting that students were attempting to employ the academic language presented in a previous lesson, and did so in ways that illustrate the connection of a term (vibration) to their experiences in, and outside of, school. Below we demonstrate the process of formalization of everyday language, as well as the process of students' trying out terminology introduced through schooling.

Students drew on everyday words throughout their discussions about the sounds produce by different strings. For example, when asked to describe what he heard when plucking the strings, Gabriel offered: "(first/shortest string goes) tick tick, (second string goes) tack tack, (third string goes) tock tock, (and the fourth) doesn't make any noise." We argue that these onomatopoeic labels are productive for various reasons. First, assigning labels that derive from observations is a sign of students establishing connections between what they are experiencing and abstract ideas. The invention of a label is a step towards exteriorizing and objectifying ideas, which plays an important role in the construction of disciplinary knowledge.

Second, referring to the strings according to the sound they made allowed other students to have access to the idea the Gabriel was trying to express. Also, onomatopoeia is a standard strategy employed in sheltered English instructional methodology. The onomatopoeic label invited students who may have been unsure about their English language skills to explain their thinking in terms that closely resemble the shared, observed phenomenon. An example is Gergö, a Hungarian student who had been in the US for less than five months and spoke little English. Referring to the strings by their sound allowed him to communicate his thinking and, in fact, label the strings: "This (first string), ting ting. And this (second string), tong tong. And (first string) small, and (second string), big and (third string) bigger." Here we see how the onomatopoeia helped Gergö talk about the link he recognized between the strings' length and the sounds' pitch. We argue that Gabriel's and Gergö's contributions were possible because of the affordances of third space; the learning environment allowed both students to bring elements of their individual everyday languages into a conversation about the physical process. It is often the case that the teacher quickly corrects students, in efforts of encouraging them to adopt and use the academic terminology even before they are ready for it.

Students also used other familiar terms such as "loose" and "hard" to describe the mechanisms that produced different sounds on the strings. Again, use of everyday terms such as "loose" and "hard" provided an entry point for students who had not yet made sense of technical terms such as "tension." For example, when trying to explain why the third string sounded "toong toong," Brian offered: "It's making it toong toong because (string) is kind of loose. (Second string) is more looser than (first string). And (third string), is more looser than (second string). That's why it's making a lower sound." In this excerpt Brian uses the everyday term loose while appropriating Gabriel and Gergö's use of "toong toong" in order to present his idea about the relationship between the strings' tension and the sounds' pitch. In this example we see the spread of the onomatopoeic terminology in a sort of classroom formalization of the terms. This made it possible for students to keep track of three rather sophisticated and critical elements of the discussion: references to specific strings, the observed differences in sound, and proposed mechanisms responsible for these differences. Moreover, as the transcript shows, talking about his idea in everyday terms ("loose") allowed Brian to propose a mechanism that could drive the relationship between the properties of the strings and the sounds they produced.

Science activities facilitated rich discussions between ELLs. Students generated and sustained a conversation based on mechanistic reasoning about sound. Right at the beginning of the session, Gabriel offered: "it makes a louder sound, the short one," which encouraged his peers to contribute their ideas about why that was so. Moreover, students were very attentive to each other's ideas, always trying to make sense of them, and usually expanding on those explanations. For example, Mahaley, a Haitian student who had been in the country for less than three months, was alluding to the length of the strings as the reason for different loudness. While students had a hard time following her explanation, they took turns trying to makes sense of what she had said, voicing their own interpretations and rewording her statement; students frequently checked with her to see if their interpretations were accurate. In the end, Gabriel noticed Mahaley was saying the "ting ting" string was tied to a hole closer to the hooked screw, putting forth the conjecture that shorter strings make high-pitched sounds. Exchanges like this happened often, demonstrating the advantages of an emergent third space in conjunction with activities based on inquiry and argumentation. We argue that observational affordances of the physics activities facilitated and mediated a rich discussion between students about mechanisms. Moreover, students responding to their peers' conjectures, and co-constructing knowledge, is evidence how third spaces distribute authority of knowledge and language among students themselves. This is contrasted to traditional models of instruction where the authority or knowledge resides with the moderator or teacher. Further, in a situation such as third space, the authority exists within the discourse between the students rather than on the academic language of the discipline, something that is outside of the students themselves.

Finally, even though we have referred to the language used when presenting ideas about mechanisms by students as everyday, we argue that these terms became formalized throughout the discussion. Returning to the example of the onomatopoeic labels, they were originally created by Gabriel's attempt to verbalize his observations, and his peers found them helpful for talking about a particular string. As we showed above, rather than rephrasing the labels, Gergö used Gabriel's labels and consulted with him to see if he was using the correct invented terminology. The continued use of the onomatopoeic labels amounts to evidence that everyday language became formalized, and established a protocol for referring to individual strings.

All along, students were well aware of the existence of academic language, and occasionally tried to use it during the discussion. For example, at the beginning of the episode, Brian said that he thought the shortest string goes "ting ting, because it hibernates faster." Here he presents a mechanism (hibernates faster) in attempt to explain why a shorter sting would create the highest pitch. In this example, Brian has attempted to try out the language that was provided in a previous class session. Gabriel questions his choice of words, "Hibernates?" and prompts Brian to correct, saying "Vibernates;" yet another attempt at utilizing the formal terminology provided by schooling. Finally, the students realize that the term is "vibrate," which is not a part of their prior experience, and until the experiment with the ruler, had no place in their discourse. Here we see how experience mediates students' connection with academic language and, more specifically, the concept of vibration.

## CONCLUSIONS AND IMPLICATIONS

Science activities are powerful experiences through which emergent third spaces can foster students' development of language and conceptual understanding. Based on the evidence presented above, we claim that students' use of everyday language was productive. Relying on socioculturally organized discourse patterns provided students the opportunity to externalize their thinking, which in turn helped them further their understanding. Additionally, the presence of familiar language in these conversations gave access to students who may not have felt unsure about their perceived level of understanding and/or language skills. The shared, observational affordances of the physical experiment facilitated discussions between students about their ideas, as well as encouraged students to make sense of their peer's question.

The discussion provided evidence of how common and invented terminology can become formalized, in a process resembling the creation of scientific discourse conventions. What originally were experiential adjectives became shared, meaningful terms to which all students had access. As it is hypothesized in third spaces, students became comfortable using the invented terminology, allowing them to express themselves and construct knowledge freely, and even to test academic language that was introduced through schooling.

These findings have important pedagogical and research implications. While preliminary, our data suggests the design of learning environments that depend on students' culturally mediated resources can be very effective in including students, and in the process of sense-making. Additionally, these findings highlight the importance of physics, and other observable sciences, in contexts focused on improving English language skills. While linguists' insights into language development are important, scientists' perspective on the connection between language and shared construction of knowledge is also crucial.

## ACKNOWLEDGMENTS

We would like to thank Brian Gravel, Ann Rosebery, and Beth Warren for their feedback and continued support. Also, we are very grateful to Hannah and all of her third graders.